\documentclass[apj]{emulateapj}
\usepackage{apjfonts,graphicx} 
\usepackage{enumerate}

\slugcomment{Submitted to ApJL on 9 March 2015; accepted on 10 April 2015}

\shorttitle{2014 ALMA Long Baseline Campaign}
\shortauthors{ALMA Partnership et al.}

\begin{document}

\title{An Overview of the 2014 ALMA Long Baseline Campaign}

\author{
ALMA Partnership,
E. B. Fomalont\altaffilmark{1,2},
C. Vlahakis\altaffilmark{1,3},
S. Corder\altaffilmark{1,2},
A. Remijan\altaffilmark{1,2},
D. Barkats\altaffilmark{1,3},
R. Lucas\altaffilmark{4},
T. R. Hunter\altaffilmark{2},
C. L. Brogan\altaffilmark{2},
Y. Asaki\altaffilmark{5,6},
S. Matsushita\altaffilmark{7}
% ...
}

\altaffiltext{1}
{Joint ALMA Observatory, Alonso de C\'ordova 3107, Vitacura, Santiago, Chile}

\altaffiltext{2}
{National Radio Astronomy Observatory, 520 Edgemont Rd, Charlottesville, VA, 22903, USA}

\altaffiltext{3}
{European Southern Observatory, Alonso de C\'ordova 3107, Vitacura, Santiago, Chile}

\altaffiltext{4}
{Institut de Plan\'etologie et d'Astrophysique de Grenoble (UMR 5274), BP 53, 38041, Grenoble Cedex 9, France}

\altaffiltext{5}
{National Astronomical Observatory of Japan, 2-21-1 Osawa, Mitaka, Tokyo 181-8588, Japan}

\altaffiltext{6}
{Institute of Space and Astronautical Science (ISAS), Japan Aerospace Exploration Agency (JAXA), 3-1-1 Yoshinodai, Chuo-ku, Sagamihara, Kanagawa 252-5210 Japan}

\altaffiltext{7}
{Institute of Astronomy and Astrophysics, Academia Sinica, P.O. Box 23-141, Taipei 106, Taiwan}

% ...

\email{efomalon@nrao.edu}

\begin{abstract}
A major goal of the Atacama Large Millimeter/submillimeter Array (ALMA) 
is to make accurate images with resolutions of tens of
milliarcseconds, which at submillimeter (submm) wavelengths requires baselines up to $\sim$15~km.  To
develop and test this capability, a Long Baseline Campaign (LBC) was carried out from September to late November 2014, 
culminating in end-to-end
observations, calibrations, and imaging of selected Science
Verification (SV) targets.  This paper presents an overview of the campaign
and its main results, including an investigation of the short-term
coherence properties and systematic phase errors
over the long baselines at the ALMA site, a summary of the SV
targets and observations, and recommendations
for science observing strategies at long baselines. Deep ALMA images of the quasar 
3C138 at 97 and 241 GHz are also compared to VLA 43 GHz results, demonstrating an
agreement at a level of a few percent. 
As a result of the extensive program of LBC testing, the highly successful SV imaging at long baselines 
achieved angular resolutions as fine as 19~mas at $\sim$350~GHz. Observing with ALMA on baselines of up to 15~km is now possible, and opens up new parameter space for submm astronomy.
\end{abstract}


\section{Introduction}\label{Sec:int}
The Atacama Large Millimeter/submillimeter Array (ALMA) is a 
millimeter/submillimeter (mm/submm) interferometer
located in the Atacama desert of northern Chile at an elevation of
about 5000~m above sea level.  The high-altitude, dry
site provides excellent atmospheric transmission over the frequency
range 85 GHz to 900~GHz \citep{MA1999}. ALMA is currently in its third year of science operations 
and was formally inaugurated in 2013 March.  Until now, science observations have used 
configurations with baselines from
100~m to $\sim$1.5~km, with some limited testing of a $\sim$3-km baseline in
2013 \citep{AS2014,MA2014}.

To test the highest angular resolution capability of ALMA using baseline lengths of up to
$\sim$15~km at selected frequencies, the three-month period from 2014 September to November was 
dedicated to carrying out 
the 2014 ALMA Long Baseline Campaign (LBC)\footnote{The LBC was led by the Extension and Optimization of Capabilities (EOC) team, which
includes members from the Joint ALMA Observatory (JAO) Department of Science 
Operations. It was a collaborative effort by an international team including members from the JAO, 
the ALMA Regional Centers, and the JAO expert visitor program.}.  
The approximate resolutions that
can be achieved with the longest baselines are 60 mas at 100 GHz, 25 mas at 250 GHz and 17 mas
at 350 GHz (but these can vary by $\sim$20\% depending on the imaging
parameters). 
The major goal of the campaign was to develop the
technical capabilities and procedures needed in order to offer ALMA long baseline array configurations for future
science observations.

This paper presents an overview of the ALMA LBC, focusing on the technical issues
affecting submm interferometry on baselines longer than a few kilometers. 
In \S\ref{sec:scope}, we describe the LBC array and campaign test strategy.
\S\ref{sec:stc} describes the effects of short-term phase
variation due to the atmosphere and 
a method for determining if conditions are sufficiently stable for imaging. 
In \S\ref{sec:spe}, we discuss the systematic phase
errors found between the calibrator and science target. 
In \S\ref{sec:SV}, an overview of Science Verification (SV) at long baselines is
given. Images and initial science results on the SV targets are presented in three accompanying papers (ALMA Partnership et al.\ 2015a,b,c).  
An illustration of the quality of the ALMA calibration and imaging is given by a comparison of preliminary ALMA SV and Very Large Array (VLA) images of 3C138 with the same resolution (Appendix A). 
In \S\ref{sec:sum}, we present conclusions drawn from the LBC and
recommendations for science observing using long baselines with ALMA.

\begin{figure}
\vspace{0.5cm}
\includegraphics[angle=00,width=8cm]{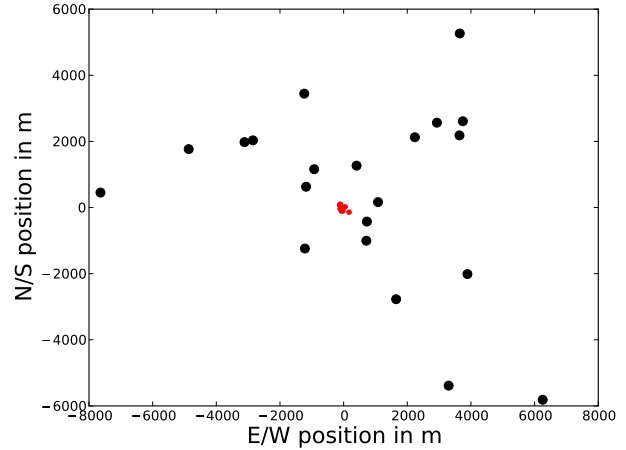}
\caption{Example LBC array configuration (in this case the array that was used for the 
3C138 Band 6 observations in Appendix A).  The black points show the nominal LBC antennas. 
The five antennas
  near the center (red points) are not part of the nominal LBC array,
  but were useful for measuring more extended emission (the number of these antennas varied; see Section~2 for details). The axis units are in meters. }
\label{Fig:LB_config}
\end{figure}

\begin{figure}
\includegraphics[angle=00,width=8cm, clip, trim = 0cm 0cm 0cm 0.6cm]{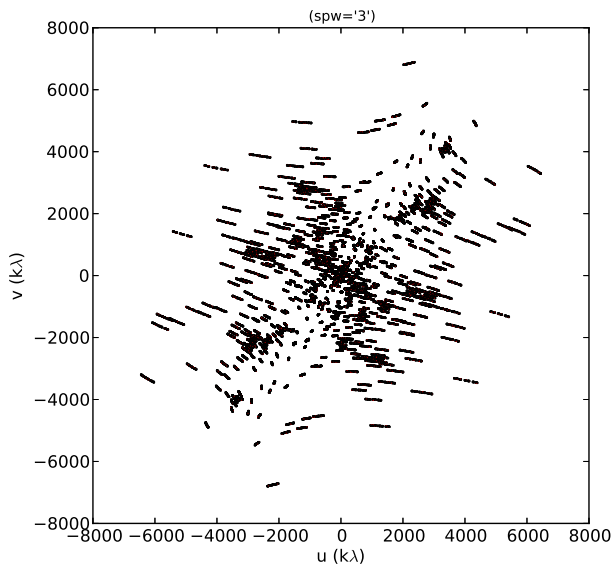}
\caption{The {\it uv} distribution for the the $\sim$1-hr 3C138 Band 6 observations.  The {\it u}-axis is the east-west spacing and
  the {\it v}-axis is the north-south spacing.  The axis units are in kilo-lambda
  (k$\lambda$, where 1000 k$\lambda$ = 1300 meters).}
\label{Fig:uv_distribution}
\end{figure}

\section{Long Baseline Campaign Overview}\label{sec:scope}
\subsection{The LBC Array}
Since many of the distant antenna pads had not been previously powered
or occupied, a coordinated effort was made from April to August 2014 to
prepare a sufficient number of antenna stations beyond 2 km from the
array center.  The configuration process began with an initial test
in late August 2014 when a single antenna was moved out to a 7~km baseline.
The nominal LBC configuration consisted of 21--23 antennas on baselines of between 
400~m and 15~km and was available from the end of 
September until mid-November 2014 (with the two longest baseline antennas being added in mid-October). 
In addition, typically 6--12 antennas were available on baselines less than 300~m that were useful for imaging the more extended sources (though since they were not part of the nominal LBC configuration, the number of these antennas on short spacings varied from day to day and with observing Band). Thus, the total number of antennas used during the campaign typically ranged from 22--36, depending on observing date and observing Band.
An example configuration
used during the campaign (in this case for the SV observations of 3C138; see Appendix A) is shown in Fig.~\ref{Fig:LB_config}.  The
resultant {\it u-v} coverage for a $\sim$1-hr observation of 3C138 with
this array is shown in Fig.~\ref{Fig:uv_distribution}.  

\subsection{LBC Test Strategy}
The normal calibration mode for ALMA observing is phase referencing
\citep{BC1995}.  Over the length of an experiment that can last for
several hours, this observing mode alternates short scans of the
science target and a nearby quasar that is used to calibrate the target data.  Hence, the outcome
of the long baseline observations depends strongly on the accuracy with which
the phase measured on the calibrator can be transferred to the target.  The
LBC concentrated on the accuracy of this
transfer by: (1) performing test observations of quasars to establish
the properties of the phase coherence of the array over long
baselines; (2) determining how to optimize observing strategy to achieve good imaging results; 
and (3) observing, calibrating, and imaging
SV targets and other test targets to demonstrate the
end-to-end capability of ALMA long baseline observations.

Key plans for the LBC testing included: (1) {\it Source stares:}
30-min observations of a single bright source to determine the
temporal phase variation statistics as a function of baseline length;
(2) {\it Short phase reference tests:} alternating observations of two
close sources to determine the accuracy of the phase transfer and
subsequent image errors; (3) {\it Go/noGo tests:} development of an
online method to determine the near real-time feasibility of long baseline
observations (Section~\ref{sec:gonogo}).  (4) {\it Cycle time tests:} phase referencing tests
with different intervals between calibrator scans; (5) {\it Baseline
  determination:} observations of many quasars distributed over the
sky for 30 to 60 min to determine antenna positions and delay
model errors; (6) {\it Weak calibrator survey:} measuring the flux
density of candidate calibrators for suitability as phase reference
sources; (7) {\it Calibrator Structures:} imaging of calibrators at
long baselines to search for significant angular sizes; and (8) {\it
  Astrometry:} phase referencing among many close quasars to measure
the long baseline source position accuracy.

Most test observations were made at 100 GHz (ALMA Band 3). The 
observed phase fluctuations are associated with variations in 
propagation time (delays) in the ALMA system or in the atmosphere, which 
are also described as path-length variations. The propagation changes 
are generally non-dispersive so that the phase fluctuations will scale 
with frequency\footnote{A useful conversion is that a path
length change of 1 mm will produce a path delay change (assuming
propagation at c) of 3.3 psec.  The 1 mm path length change will
produce a phase change of $120^\circ$ at 100 GHz (Band 3),
$300^\circ$ at 230 GHz (Band 6), and $420^\circ$ at 340 GHz (Band 7).} 
(although there are significant dispersive effects in 
the contributions due to water vapor at some frequencies above 350~GHz; these effects can be estimated).

\section{Short-term Coherence}\label{sec:stc}
Imaging using phase referencing techniques requires a
reasonably phase-stable array. Hence, an early goal of the LBC was to
determine the short-term (5 to 60 sec) phase rms properties of
ALMA over a variety of conditions. In addition to phase noise, systematic 
phase offsets between the science target and calibrator were found; 
in \S\ref{sec:spe}, we describe their origin and how they were minimized.

One of the main contributions to phase instability at mm wavelengths is the 
fluctuation of the amount of water vapor in the atmosphere.
The ALMA site was chosen for its low average water vapor content and
excellent phase stability.  Nevertheless, at baselines longer than 1~km, 
the short-term phase variations may make imaging impossible.  A
good rule of thumb is that if the rms phase variations are $\sigma$
(rad), then the approximate loss of coherence (the decrease of the
peak intensity of a point source caused by these random phase
fluctuations) is exp[($-\sigma^2/2$)] \citep{RI2003}.  For $\sigma =
30^\circ$ or $60^\circ$ the coherence is respectively 87 or 58\%.
Hence, a general guideline is that the loss of coherence is acceptable and
reasonably accurate image quality can be obtained if the rms phase fluctuations are
$<30^\circ$.

\begin{figure}
\centering
\includegraphics[angle=00,width=8cm]{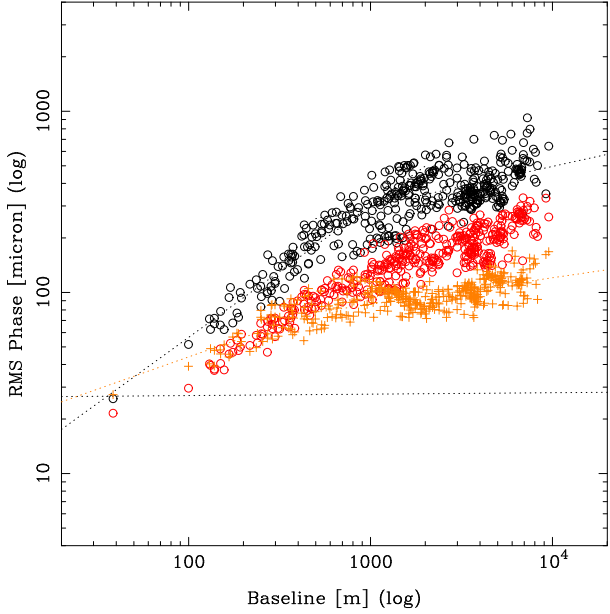}
\caption{The spatial structure function (SSF). The phase rms \protect\footnote{See footnote 44.} 
(square-root of the SSF; converted to a path length in microns) versus baseline
  length is shown for a target at three stages of reduction.  The
  experiment was 15 minutes in duration.  The black points show the
  SSF for the original visibility data.  The red
  points show the SSF points after applying the WVR correction for
  this source.  The orange points show the SSF for this source {\it
    after} phase referencing with a calibrator that is $1.3^\circ$
  away from the target with a cycle time of 20 sec. The PWV during
  this experiment was 1.44~mm with a wind speed of 7 m/sec.}
\label{Fig:spatial_structure}
\end{figure}

\begin{figure}
\vspace{0.5cm}
\includegraphics[angle=00,width=8cm]{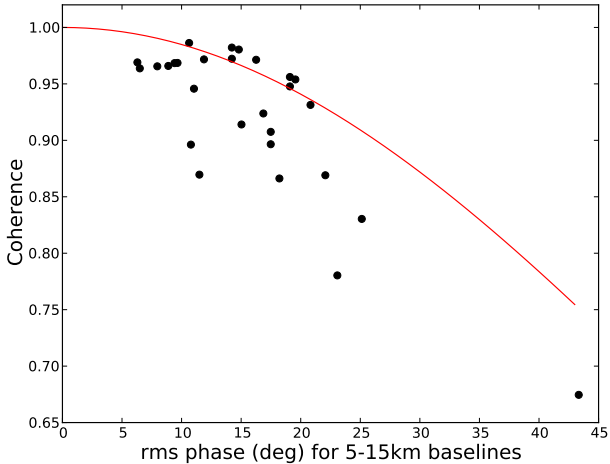}
\caption{Average phase rms for 5 to 15~km baselines versus coherence. 
  The rms phase was determined from a set of {\it Go/noGo} (see Section ~\ref{sec:gonogo}) observations that were
  followed by short phase referencing experiments.  The coherence is
  the ratio of the phase referenced image peak density divided by the
  self-calibrated image peak flux density.  The thin red line shows the
  theoretical relationship between the phase rms (radians), $\sigma$,
  and coherence, i.e. exp($-\sigma^2/2$) for a random phase distribution (see Section~\ref{sec:stc}).}
\label{Fig:phase-peak}
\end{figure}

\begin{figure*}
\centering
\vspace{0.3cm}
\includegraphics[angle=00,width=11cm,clip, trim = -2.0cm 0cm 0cm 0.8cm]{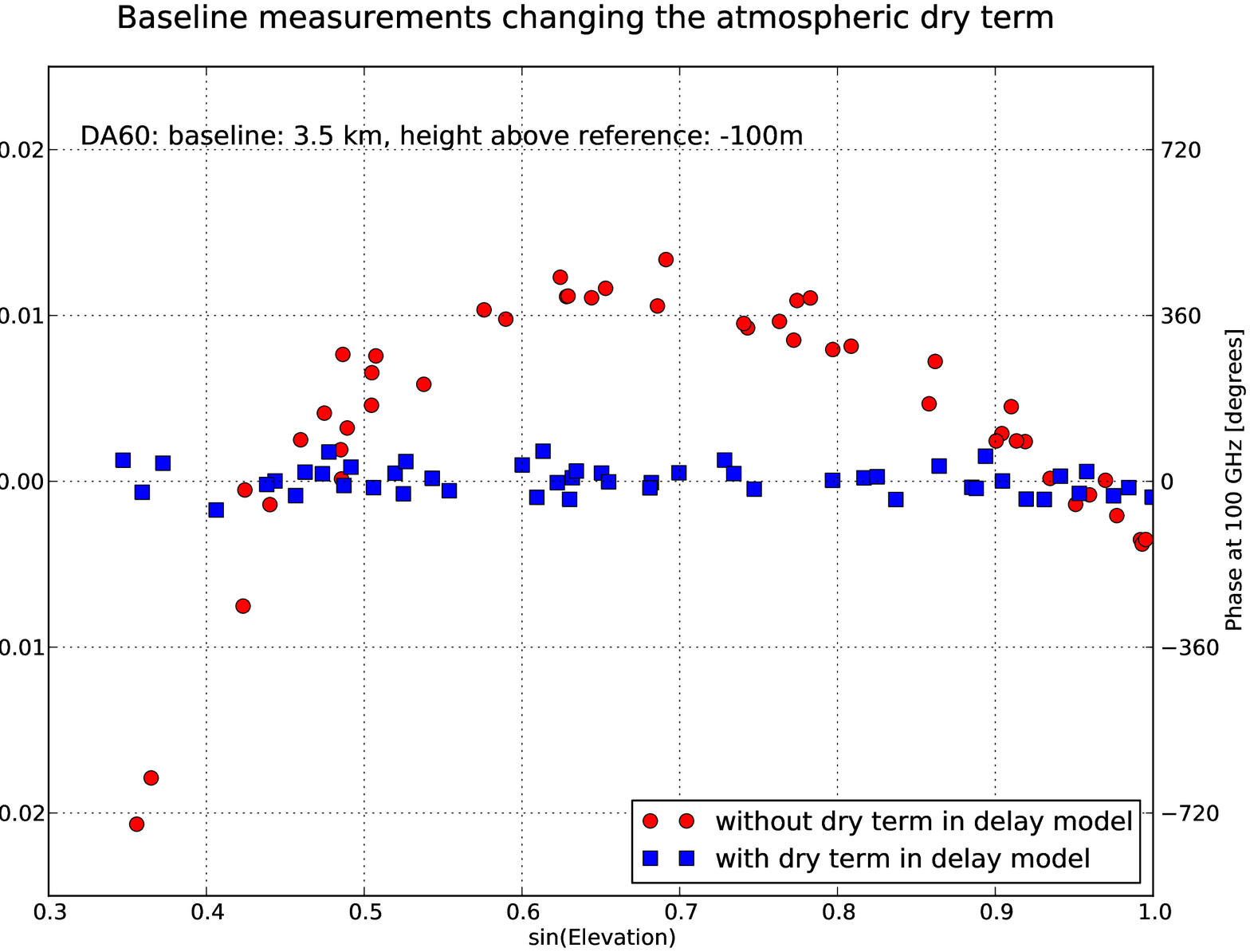}
\caption{The effect of the height difference delay term. The residual
  delay/phase for one baseline (after fitting for the best antenna positions) 
is plotted versus the sin(elevation) for 50 quasar scans that form a typical baseline observation. 
The baseline length is 3.5 km with an antenna height
  difference of 100 m.  The red points show the residual delay/phases
  that used the nominal ALMA CALC delay model (Section~\ref{sec:dm}) that assigned the
  measured pressure from the one weather station near the array center
  to {\it both} antennas.  The blue points show the residual
  delay/phase for another baseline observation in which the estimated
  pressure for each antenna was estimated using the pressure lapse
  rate from the barometer near the array center.}
\label{Fig:dry}
\end{figure*}

\subsection{WVR correction and the Spatial Structure Function}

To estimate the path variations associated with the water vapour component, 
each antenna is equipped with a Water Vapor Radiometer (WVR).  
The WVR is a multi-channel receiver system (Emrich et al. 2009) that makes continuous observations 
of the emission in the wings of the 183 GHz water line along the line of sight to the astronomical source.  
A description of this system, and of the way in which the measurements are used to estimate the variations 
in the amount of Precipitable Water Vapor (PWV)\footnote{Each mm of PWV along the line of sight will result in a path length increase of 6.5mm; \citet{TH2001}.} in the path to each antenna, is given in \citet{NI2013}. 
This WVR correction typically removes about half the short-term phase fluctuations,
and increases the proportion of time that phase referencing
observations will produce good quality images. 
Even in good conditions, however, applying a correction to the phases based on these estimates 
still leaves residual fluctuations that are much larger than the estimated errors 
(which, with clear skies and PWV < 2mm are believed to be less than 20 microns, 
although they can be much larger when clouds of ice or liquid water are present); see Figure~3.  
These residuals are thought to be mainly due to dry atmosphere (i.e. density) fluctuations 
(also see Section~\ref{sec:dryterm}).

The properties of the phase rms as a function of baseline length are
important for deciding when and how to observe at long baselines.
Fig.~\ref{Fig:spatial_structure} shows a typical relationship of the
phase rms, $\sigma$, as a function of baseline length, $b$, for a
target at three stages of analysis.  The $\sigma-b$ relationship is
called the Spatial Structure Function (SSF).  The characteristic shape is
similar for both the uncorrected data and for the WVR
corrected data, except for the decrease of the
variations by about 50\%.  For short baselines, the rms phase
increases as $\sigma\approx b^{0.83}$, indicative of a 3-D Kolmogorov
spectrum \citep{CH1999}.  The slope then decreases to a 2-D Kolmogorov spectrum with
dependence $\sigma\approx b^{0.33}$ at about 3 km, which is roughly
the scale height of phase turbulence.  This scale height is an average
of the wet atmosphere and dry atmosphere scale sizes of 1 and 5 km at 
the ALMA site.

After phase referencing, the shape of the SSF is altered, as shown by
the orange points in Fig.~\ref{Fig:spatial_structure}.  In this
example, the calibrator is only $1.3^\circ$ away from the target, 
the cycle time is 20 sec and the integration time on the calibrator is only 6 sec.
Only a small fraction of calibrators are sufficiently strong, even at Band 3, 
to provide adequate signal-to-noise for accurate phase referencing calibration 
in this short integration time. 
Even in the ideal case of a sufficiently strong calibrator, for baselines less than 1 km, there is
little decrease in the target rms after phase referencing.  However,
beyond a baseline of about 1 km, the target rms becomes less dependent on baseline length since
the phase fluctuations with scale sizes greater than 1 km are well
correlated between the target and calibrator with a 20 sec switching
cycle time.

\subsection{Go/noGo System}\label{sec:gonogo}

At the beginning of the campaign, it was hoped that the properties of
the rms phase fluctuations (both before and after WVR correction)
could be predicted from measurable weather parameters such as the
average PWV, PWV rms, wind speed, and pressure rms.  If
so, then algorithms associated with these measured conditions could be
used to indicate in advance if the phase parameters are adequate for imaging at
a specified frequency; namely, that the short-term phase rms would be
less than about $30^\circ$ for the longer baselines. 
This presumption, however, turned out to be
not always true. 

A direct method to determine the current ALMA phase rms is from a
short observation of a strong source.  A simple observing procedure
called {\it Go/noGo} was developed, consisting of a 2-min observation
of a strong quasar at Band 3, followed by online data
analysis that rapidly determines the SSF with the WVR
correction applied. 
To confirm that the {\it Go/noGo} structure function phase rms (averaged
over many baselines between 5 to 15~km) is well correlated with
phase referencing image quality, many {\it Go/noGo} observations that were carried out during the
LBC were followed by short reference observations of calibrator-target pairs, with a typical $3.5^\circ$ separation and
cycle time of 60 sec.  The plot of the {\it Go/noGo} rms phase versus image
coherence from the phase referencing experiment is shown in
Fig.~\ref{Fig:phase-peak}.  This demonstrates that the
target image coherence is reasonably well correlated with the rms
phase at the longer baselines of the calibrator.  The reason for the
somewhat lower image coherence than expected from the rms phase variations
are discussed in \S\ref{sec:spe}.

\section{Systematic Phase Errors}\label{sec:spe}
In addition to the stochastic-like phase variations between the
calibrator and target described in Section~\ref{sec:stc}, there were systematic
antenna-based phase offsets between the calibrator and science target that
persisted on timescales of many minutes to hours. These were found to be caused mostly by errors in
the correlator delay model.  The offsets were found to scale roughly as the
calibrator-target separation, but were nearly unaffected by the cycle
time.  Such systematic offsets can have serious impact on the target
image quality because they are persistant and
produce image artifacts (e.g. large side-lobes and spurious faint
components), in addition to the blurring of the target image that is 
associated with short-term phase fluctuations.

\begin{deluxetable*}{ccccccc}
\tabletypesize{\footnotesize}
\tablecaption{Long Baseline Astrometric Results}
\tablewidth{0pt}
\tablehead{
\colhead{Source} & \colhead{J0538-4405\tablenotemark{a}} & \colhead{J0519-4546} & \colhead{J0455-4615} & \colhead{J0522-3627} 
}
\startdata
Sep (Deg)  &      0.0                 &     3.8      &     7.9    &      8.2              \\
RMS (mas)\tablenotemark{b}  &      0.13              &    0.20      &    0.35      &     0.08            \\
& \\
DATE       & \multicolumn{4}{c}{RA, DEC offset (mas)}                                               & PWV(mm)  & ELEV(deg)  \\ 
Sep 22     & --0.45, +0.15       & --0.44, --2.20  & +11.0, --2.96  & +0.48, +1.82 &   0.7  &  66   \\
Oct 03     & --0.09, --0.10    & +0.28, --4.57   & +4.03, --8.14  & +1.32, --2.99&   2.7  &  71   \\
Oct 14     & +0.06, +0.08      & --2.69, +0.25   &      ----            & +0.74, --3.49 &   0.6  &  55   \\
Oct 14     & --0.00, --0.03   & +0.54, +1.27   & +8.81, +3.84    & --4.00, --1.70 &   0.9  &  71   \\ 
Nov 04     & --0.03, --0.02   & --0.21, +2.12   & +7.08, +4.80    & --4.60, +0.60 &   0.9  &  48   \\
Nov 17     & --0.01, +0.05   & --2.72, --1.24 & --6.56, --1.96   & --9.02, --0.17 &   1.3  &  66   \\
Mean       & --0.09, +0.02   & --0.87, --0.73  & --4.87, --0.88   & --2.51, --0.99  &&             \\
STD (Mean) &  0.08,  0.03    &  0.54,  0.92       &  2.75,  2.15       &  1.15,  0.78     &&      \\
\label{tab:astrometry}
\enddata
\tablenotetext{a}{Phase reference source.}
\tablenotetext{b}{Theoretical RMS, defined as the angular resolution divided by the theoretical SNR, where the latter is derived from the peak flux density of the source divided by the expected image rms noise level.}
\tablecomments{Details are given in Section~\ref{sec:astrometry}.}
\end{deluxetable*}

\subsection {The Delay Model}\label{sec:dm}
The signals from all antennas must be combined precisely in phase at
correlation to obtain accurate visibility phases.  A critical part of
the ALMA online control software, called the delay server, calculates
the expected relative delay of the signals between each antenna from
the ALMA array parameters \citep{MA2008}.  
If the delay model (DM; which is calculated using the 
CALC\footnote{\url{http://lacerta.gsfc.nasa.gov/mk5/help/calc\_01.txt}} third-party software) 
is accurate, the visibility phase for any point-like quasar
with known position should be constant with time and independent of
the quasar's position in the sky.

An important part of the DM is the estimate of the differential
tropospheric delay between each antenna from the source.  As described
above, the wet delay component is calculated from the 183 GHz emission
assuming a model temperature profile, and is included in the DM using
the WVR measurement.  The zenith dry air delay $\tau_i$ above antenna
$i$ is accurately given by $\tau_i\approx 0.228P_i$ where $P_i$ is the
dry pressure in mbars at the antenna \citep{TH2001}.  For an
observation of a target at elevation $e$, the CALC model delay is
$(\tau_i)/sin(e)$.  Given that only one weather station near the array
center had so far been available at the time of the LBC\footnote{Installation and testing of several additional weather stations distributed over the array is planned for the end of 2015.}   at ALMA, the
estimate of the dry air delay at each antenna is not as accurate
as desired. This inaccuracy results in antenna-based phase offsets
that differ between a calibrator and science target and hence produce
relatively constant phase offsets between them.

\subsection {Measurement of Delay Model Errors}\label{sec:dryterm}

The presence of DM errors was suspected from the baseline observations
that consisted of about 50 to 100 ten-second quasar observations distributed
over the sky\footnote{\url{http://legacy.nrao.edu/alma/memos/html-memos/alma503/memo503.pdf}}. 
Many such observations have been made in order to determine the accurate relative positions of the
antennas which are frequently moved from one antenna pad to another as the ALMA
configuration changes.  The a priori antenna positions are usually
more than 1~mm in error, so the baseline observations provide the
data needed to update antenna positions, generally to an accuracy of about 50
microns.  Over a few years, it was found that the measured position
changes of fixed antennas between baseline calibration observations, separated by several
hours to a few weeks, were often larger than 100~microns and sometimes
well over 1~mm for unmoved antennas that were more than 1~km from the
array center.

These apparent antenna position changes were traced to the implementation
of the dry air delay term in the CALC DM.  Fig.~\ref{Fig:dry}
illustrates the results of an experiment on 2014 September 16 with two
30~min baseline observations which confirmed the DM error for a 3.5 km
baseline with a height difference of 100 m between the two antennas.
One experiment used the DM in which the pressure at each antenna was
set equal to that measured by the one sensor.  After fitting for the
best baseline, the residual fit, shown by the red points, contains a
large residual phase versus elevation term.  In the subsequent
experiments, the pressure at each antenna was estimated using the
approximate pressure lapse rate.  After the best baseline fit to the
data, the residual phase versus elevation is flat.  Since some
antennas in the long baseline array have a height difference from the array
center of over 200~m, even larger systematic phase errors could be
encountered.

Without a reasonable pressure estimate for each antenna, the target and phase
will have a systematic offset that will only slowly change.  For
example, the residual phase between a calibrator at elevation
$55^\circ$ and a target at elevation $60^\circ$ is about $110^\circ$; 
this phase offset is not removed by the phase referencing.  After the
September demonstration of the issue, the ALMA DM was updated
to include an estimate of the pressure at all antennas using the lapse pressure rate
and the height of the current single pressure monitor (as noted in Section~\ref{sec:dm}, additional pressure monitors distributed across the array will be available in future).  This height-delay
compensation is also used at the VLA 
\citep{FO1999}.

Even after the correction of the antenna height delay differences,
additional baseline observations during the last part of the LBC still
showed apparent antenna position offsets of about 1-5 mm for most
antennas 5 to 10~km from the center, which scaled roughly
with distance from the array center.  These apparent antenna position changes are
consistent with the un-modeled pressure changes expected over the 15 km
region of the Chajnantor plateau.  However, by using a calibrator close to the target this effect is minimized; this requires a larger catalog of potential calibrators (Section~\ref{sec:weakcal}).

Additional observational techniques can be employed to model the 
dry term delay residuals.  For example, Very Long Baseline Array (VLBA) observations often 
include a short baseline-type observation (20 sources in 20 min) to 
determine the residual zenith path delay over each antenna \citep{MR2014}. 
Such options may be explored for future work.

\begin{deluxetable*}{lcccccllll}
\setlength\tabcolsep{1pt}
\tabletypesize{\scriptsize}
\tablecaption{ALMA Long Baseline Science Verification Targets\label{tab:SV}}
\tablewidth{0pt}
\tablecolumns{10}  
\tablehead{
\colhead{Target} & \colhead{Coordinates\tablenotemark{a} } & \colhead{Band\tablenotemark{b} } & \colhead{Scope\tablenotemark{c} } & \colhead{N$_{ant}$\tablenotemark{d} }  & \colhead{N$_{ex}$\tablenotemark{e} } & \colhead{t$_{ON}$\tablenotemark{f} } & \colhead{Freq.\tablenotemark{g} } &  \colhead{Obs. Date\tablenotemark{h} } & \colhead{Id.\tablenotemark{i} } 
}
\startdata
Juno & ephemeris target & 6 & cont., ephemeris & 30-33  & 5 & 0.3$^{\ast}$ &   224.0, 226.0, 240.0, 242.0 & 10/19  & 13\\
Mira & 02$^{h}$19$^{m}$20$^{s}$.79 -02$^{\degr}$58$^{\arcmin}$39$^{\arcsec}$.5 & 3 & SiO, cont. & 31-33  & 3 & 1.5 &  88.2, 98.2, 100.2, 86.8,  & 10/17-10/25 & 14\\
&&&&&&&86.2, 85.6, 85.7 && \\
        &  & 6 & SiO, cont. & 35-36  & 3 & 1.0 & 229.6, 214.4, 214.1, 215.6,  & 10/29-11/01   & 14\\
&&&&&&&217.1, 232.7, 231.9&& \\
HL~Tau & 04$^{h}$31$^{m}$38$^{s}$.45 +18$^{\degr}$13$^{\arcmin}$59$^{\arcsec}$.0 & 3 & CO, CN, cont. & 32-35 & 7 & 3.2& 102.9,101.1,115.3,113.5 & 10/28-11/14  & 15 \\
        &  & 3 &  HCN, HCO$+$, cont. & 33-35  &  7 & 3.5 & 90.8,100.8,102.8,88.6,89.2 & 10/14-11/13  & 15 \\
       &  & 6 & cont. & 28-36  & 9 & 4.7 & 224.0,226.0,240.0,242.0 & 10/24-10/31 & 15 \\
       &  & 7 & cont. & 27-36  & 10 & 5.1 & 336.5,338.4,348.5,350.5 & 10/30-11/06  & 15 \\
3C~138  & 05$^{h}$21$^{m}$09$^{s}$.9 +16$^{\degr}$38$^{\arcmin}$22$^{\arcsec}$  & 3 & cont., polarization & 27-30  & 6 & 2.0 & 90.5, 92.5, 102.5, 104.0 &11/10-11/19  & ...\\
       &   & 6 & cont. & 29-31   & 5 & 1.6 & 224.0, 226.0, 240.0, 242.0 & 11/09-11/14  & ...\\
SDP.81 & 09$^{h}$03$^{m}$11$^{s}$.61 $+$00$^{\degr}$39$^{\arcmin}$06$^{\arcsec}$.7  & 4 & CO, cont. & 22-27  & 12 & 5.9 & 144.6,154.7,156.4,142.7 & 10/21-11/11  & 16\\
      &  & 6 & CO, H$_{2}$O, cont. & 30-36  & 9 & 4.4 & 228.0, 230.0, 243.0, 244.5 & 10/12-11/09  & 16\\
      &  & 7 & CO, cont. & 31-36   & 11 & 5.6 & 282.9, 294.9, 296.9, 284.9 & 10/30-11/04  & 16\\
\enddata
\tablenotetext{a}{Coordinates of the phase center (J2000)}
\tablenotetext{b}{ALMA Bands. Bands 3, 4, 6, \& 7 correspond to frequencies of approximately 100 GHz, 140 GHz, 230 GHz \& 340 GHz, respectively.}
\tablenotetext{c}{Scope and aim of the observations. These include spectral line and/or continuum imaging at high angular resolution, plus polarization and ephemeris targets.}
\tablenotetext{d}{Number of antennas in the array for each execution. Typically, between 1-5 of the total number of antennas were flagged for a given execution. Note that the number and configuration of the antennas on very short spacings varied from day to day (see Section~\ref{sec:scope}). The number of antennas also varied with observing Band, with the fewest antennas available in Band 4 (due to fewer antennas with Band 4 receivers available during the LBC). }
\tablenotetext{e}{Total number of executions of the scheduling block.}
\tablenotetext{f}{Total effective integration time on source (i.e., after flagging), in hours. $^{\ast}$For specific details of Juno, see ALMA Partnership et al. (2015a).}
\tablenotetext{g}{Mean center frequency of each spectral window (spw) in GHz. 
Channel widths were 15.6~MHz for continuum windows, 2.0~GHz bandwidth.
Channel widths varied for spectral line windows. For Mira, they were 61-122~kHz, 0.059-0.117~GHz bandwidth. For HL~Tau, they were 61 kHz,  0.117 or 0.243~GHz bandwidth. For SDP.81, they were 0.488-1.953~GHz, 1.875~GHz bandwidth.}
\tablenotetext{h}{Range of dates of the observations.}
\tablenotetext{i}{The project code identifier of the dataset can be obtained by replacing ``XX'' in  \dataset{ADS/JAO.ALMA\#2011.0.000XX.SV} with the number in this column.}
\tablecomments{Further details of the Juno, HL~Tau, and SDP.81 observations and results are given in three accompanying papers (ALMA Partnership et al.\ 2015a,b,c). The data is publicly available from the ALMA Science Portal\footnote{http://www.almascience.org}. }
\end{deluxetable*}

\subsection{The Weak Calibrator Survey and Calibrator Structure}\label{sec:weakcal}

To facilitate an optimal calibrator choice for a science target, most
observatories support a source catalog that contains information about candidate calibrators.  The ALMA
calibrator catalog\footnote{\url{https://science.nrao.edu/facilities/alma/aboutALMA/Technology/ALMA\_Memo\_Series/alma599/memo599.pdf}, \url{http://www.eso.org/sci/publications/messenger/archive/no.155-mar14-2014/messenger-no155.pdf}
  page 19} in September 2014 contained 700 entries of quasars
with positional accuracy $<2$ mas from Very Long Baseline Interferometry (VLBI) observations and with a
100 GHz flux density $>25$~mJy.  Over the ALMA sky between $-90^\circ$
and $+45^\circ$ declinations, the mean angular distance of an ALMA
catalog entry from a random target is $3.5^\circ$ with a 25\% chance
that the closest calibrator is $>5^\circ$ away.  The number of
suitable calibrators in the catalog, especially for the long baseline
observations, therefore needed to be substantially increased. 
To this end, a survey of weak calibrators was initiated in mid-September to
observe candidate sources from the AT20G \cite{MA2011} and VLA
calibrator
catalogs \footnote{\url{http://www.aoc.nrao.edu/$\sim$gtaylor/csource.html}} to
determine their flux densities at 100 GHz. This list of 4200 candidate sources was compiled from sources potentially
stronger than 25 mJy at 100 GHz, and observations prioritized the $\sim$3000
sources with VLBI positions\footnote{\url{http://astrogeo.org/vlbi/solutions/rfc/atmos\_2014d}} having a positional
accuracy of $<2$ mas. 
Sources as faint at 10~mJy at Band 3 may potentially be used as phase 
calibrators, but finding the faintest acceptable calibrators will probably require future 
targeted searches around a source. \footnote{\url{https://science.nrao.edu/facilities/alma/aboutALMA/Technology/ALMA\_Memo\_Series/alma493/memo493.pdf/}}

About 20 of the brightest ALMA calibrators were also imaged with the LBC 
array to determine if they were resolved at the longer
baselines.  Since most of the sources have been previously imaged using
VLBI baselines of 5000 km at cm-wavelengths and found to be less than
about 5 mas in angular size, it was expected that these calibrators
would be nearly unresolved sources at ALMA long baseline resolutions.
Two of the 20 sources, however, had 
faint inner jets whose brightness was a few percent of the bright core point 
component, but this structure has little effect on their use as 
calibrators of amplitude and phase on long baselines.  
A few of the brighter calibrators were already known to have 
large arcsec-scale structure (J0522-3627 and 3C273); this also 
has no significant effect on their use as long baseline
calibrators.

\subsection{Astrometric Accuracy}\label{sec:astrometry}
During the campaign, many hour-long experiments, cycling among three or
four quasars within a radius of $10^\circ$, were carried out.  All of the
quasars have an a priori position accuracy of $<0.3$ mas, and were
observed sequentially with 1-min scans at Band 3.  Using one of the
quasars as the phase reference calibrator, images of the other quasars
were obtained and the positional offset for each source was determined
by the displacement of the quasar peak from the center of the image.
Uniform weighting with only spacings longer than 1 km were used to
obtain the highest resolution and most accurate positions.

The results for the same quartet of quasars observed six
times over the LBC are given in
Table~\ref{tab:astrometry}.  The source J0538-4405 is
the phase reference source, so its position should be close to zero.
The separation of the sources from J0538-4405 in degrees and the
theoretical positional rms error in mas are listed in the first two rows. 
The subsequent rows then give the R.A. and decl. offset for
each source for each of the six observations, with the mean positional
offset and the standard deviation of the mean at the bottom.  
The results show that the positional offsets of the three target sources
are significantly larger than those expected from the image noise
alone (typically $\sim$1-5~mas).  The source, J0519-4546, closest to the phase reference source,
shows the smallest systematic offset ($\sim$0.8~mas).  The other two sources, one to
the east and one to the south of J0538-4405, have larger offsets.
This relationship is consistent with that produced by the relatively
systematic atmospheric delay model errors discussed in Section~\ref{sec:dryterm}. 
In future, it will be possible to use the apparent positional
offsets of three calibrators to determine more information about the
delay model error over the array, and then remove the errors to obtain
more accurate positions of the calibrators.  Such multi-calibrator
observations and analyses have proved successful with the VLBA for
significantly improving the astrometric precision \citep{FO2002} and
are now being tested for ALMA.

The nominal astrometric accuracy from the LBC tests, given by
the average rms in Table 1 for the three sources, is an rms
positional error of $\sim$1.5 mas.  This is for an average calibrator-target separation
of $\sim6^\circ$, with an observing period of one hour, with a
maximum baseline of 12~km.  Given a sufficiently strong point source, this accuracy is independent of observing frequency. The predicted ALMA astrometric accuracy is
$\sim$0.18 mas (Lestrade 2008), assuming the use of WVR
corrections and a typical calibrator-target separation of $6^\circ$ (which is within
the range used in the LBC).  However, this predicted
value assumes that the pressure measurement at each
antenna would be accurate to  $\pm 2$~mbars.  As discussed in Section 4,
with the availability of only one weather station during the LBC, the
inferred pressure for antennas many kilometers from the pressure sensors,
 using a simple plane-parallel atmosphere model and lapse rate, could be in
error by tens of mbars.  This produces a systematic phase error
between calibrator and target and is likely the major cause of the
poorer than expected astrometric accuracy observed during the LBC. 
It is expected that the addition, in late 2015, of more weather stations distributed over the array will improve the astrometric accuracy.

\section{Science Verification}\label{sec:SV}
Science Verification (SV) is the process of fully testing observing modes
expected to be available for science observing by making end-to-end observations (e.g. execution
of scheduling blocks, calibration, and imaging) of a small number of
selected astronomical objects. The aim is to demonstrate that ALMA is
capable of producing data of the quality required for scientific
analysis so that the observing mode
can be offered for future science observations.  To demonstrate ALMA's high
angular resolution capability, during the LBC we carried out SV
observations of five targets chosen from a broad range of science
areas (Table~\ref{tab:SV}). The aim was to produce high fidelity, high resolution, images of
continuum and spectral line emission using the LBC array.

The SV targets were chosen primarily based on their suitability for
demonstrating the long baseline capability e.g., having fine-scale angular structure,
being less than two arcsec in size, being observable at night-time during the campaign period, and, where possible, having previous observations with other
telescopes.  The targets were: Juno, an asymmetric asteroid
with a 7.2-hour rotation period; Mira, a well-studied AGB star that is
the prototypical Mira variable; HL Tau, a young star with a
circumstellar disk; 3C138, a strongly polarized extended quasar; and 
SDP.81, a high-$z$ ($z$=3.042), gravitationally lensed, submm galaxy. Details of the targets and observations are given in Table~\ref{tab:SV} and the data are publicly available from the ALMA Science Portal\footnote{\url{http://www.almascience.org}}. Examples of the SV imaging results are given in three accompanying
papers on targets HL~Tau, Juno, and SDP.81 (ALMA Partnership et al. 2015a,b,c). Angular resolutions achieved were as fine as 19~mas (Band 7; 344~GHz; ALMA Partnership et al. 2015b). 
In Appendix~\ref{sec:3c138}, we compare preliminary ALMA results on 3C138 with
a 43~GHz VLA image.  
Details of the imaging of the SV targets, including important lessons learned, are described in a CASA guide page\footnote{\url{http://casaguides.nrao.edu/index.php?title=ALMA2014\_LBC\_SVDATA}}. 
Specific comments concerning the use of self-calibration to improve image quality are given in Appendix~\ref{sec:selfcal}.

\section{Conclusions}\label{sec:sum}

The 2014 ALMA Long Baseline Campaign achieved an increase of a factor of $\sim$6 in maximum baseline length ($\sim$15~km) compared to previous test observations and a factor of $\sim$10 increase compared to previous ALMA science observations (a factor of $\sim$100 smaller beam area). Further testing will be carried out in future
to extend the maximum baseline to $>$15.0 km and to higher frequencies.

Some specific results drawn from the campaign are as follows.
\begin{itemize}

\item Phase referencing observations should only be made when the short-term 
phase rms is $<30^\circ$, unless the target source is relatively compact and 
strong enough for self-calibration.  This applies to all ALMA observations, 
regardless of maximum baseline length or frequency.

\item Under clear skies, the WVR correction typically improves the phase noise by a factor of $\sim$2. 
The remaining phase fluctuations are thought to be mostly due to dry atmosphere variations.

\item The prediction of short-term phase variability cannot be made reliably using 
ground-based measurements.  Short observations of a strong source are the most 
reliable methods to determine phase conditions, as described in the {\it Go/noGo} procedure.

\item Systematic phase differences between calibrator and science target are 
dominated mainly by the lack of an accurate dry atmospheric delay 
model.  Additional pressure sensors distributed across the ALMA 
array will in future improve the models. 

\item The phase referencing cycle time recommended for long baseline 
observations is 60 to 90 sec between calibrator observations. Shorter times do not improve 
significantly the image quality unless a calibrator $<1.5^\circ$ 
from the target is sufficiently strong that it can be detected with a 6-sec integration.

\item The survey of weak calibrators will continue in order to increase the number of sources in the catalog and increase sky coverage. Alternative calibrator observing strategy may be needed in future in order to find the faintest acceptable calibrators.

\item The integration time on source may in many cases be driven by the 
time needed to obtain sufficient {\it u-v} coverage, rather than that needed 
to reach a specified rms.  In future, detailed simulations may be 
needed to investigate this.

\item More sophisticated methods of self-calibration may be needed for extended 
sources where the SNR on the longer baselines drops 
below that needed for self-calibration using one reference antenna.
\end{itemize}

As a result of the extensive program of testing during the LBC, Science Verification at long baselines was highly successful, resulting in angular resolutions as fine as 19~mas. Initial science results on the SV data are presented in ALMA Partnership et al. (2015a,b,c). The LBC has allowed long baseline (up to $\sim$15~km) antenna configurations to be made available for science observations. This fulfils a major goal of ALMA to accurately image sources at
mm and submm wavelengths with resolutions of tens of
milliarcseconds, and, together with ALMA's high sensitivity, opens up new parameter space for submm astronomy.

\appendix

\section{ALMA Observations of 3C138}\label{sec:3c138}
The source 3C138=J0521+1638 is a compact steep spectrum quasar with
$m_v=18.84$ and a redshift of 0.759 \citep{CO1997}.  Its angular size
is about $0.4''$ and consists of a radio core, with a strong jet/lobe
to the east and a weaker counter-lobe to the west.  The integrated
source linear polarization is 10\% and its total flux density is
relatively stable.  

The source 3C138 was chosen as an SV target because
its angular size and small-scale structure are ideal for imaging with
the ALMA long baselines, it is a highly polarized target, and the ALMA resolutions
at Band 3 and Band 6 with a 5 to 15 km baseline array are comparable
to that of the VLA 35 km baseline array at 43~GHz.
Thus, a detailed comparison of the images made with different
arrays can be made. For the other LBC SV targets, the ALMA
resolution and sensitivity far exceed those of other arrays so any
detailed comparison cannot be made.  Hence, the discussion here will
concentrate on ALMA--VLA comparison, rather than any astrophysical
interpretations.  The analysis of the complete set of 3C138
ALMA observations (with full polarization) is in progress. Here, we present preliminary results 
\footnote{The ALMA Band 3 \& 6 avergage frequencies for the initial results presented here are respectively 97 and 241~GHz; only the upper sideband of the Band 6 data was used.}. 

The ALMA observation parameters for Bands 3 and 6 are listed in Table
2.  The VLA observations at 43 GHz were made on February 16, 2014 in the
A-configuration, and the integration time on 3C138 was 45 min.  The
VLA observations used J0530+1331 as the phase calibrator, while the ALMA
observations used J0510+1800, both of which are within $4^\circ$ of
3C138.  The flux density scale for ALMA was based on the derived flux
density of 1.20 Jy and 0.97 Jy (10\% uncertainty) for J0510+1800 at 97
and 241 GHz, respectively.  For the VLA, the source 3C48 was used for
the flux density scale.  The phase referencing cycle time was 95 sec
for ALMA and 90 sec for the VLA.

The standard phase referencing calibration, editing, imaging, and
self-calibration for the ALMA and VLA data was carried out using the {\it
  obit}\footnote{Note that the ALMA data could have been processed in
  CASA 4.2.2 or higher, but was done in {\em obit} for consistency
  with the previously reduced VLA data.} software package
\citep{CO2008}.  Since the structure of 3C138 is dominated by a small
component, the self-calibration process was straight-forward.  In
order to compare the images at the three frequencies at the same
resolution (91$\times$51 mas in P.A. $-13^\circ$), each data set was weighted
to include approximately the same range of spacings for each image,
and then convolved with the above Gaussian beam size.

\begin{figure}
\includegraphics[angle=00,width=8cm]{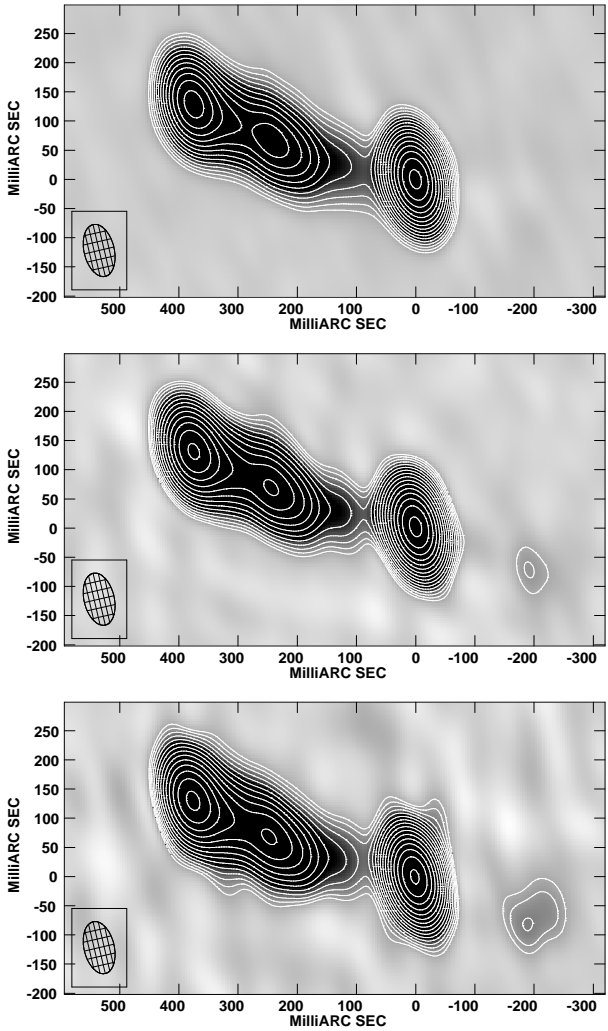}
\caption{Images of 3C138 with 91 x 51 mas resolution and P.A. $13^\circ$ (as
  shown by the cross-hatched ellipse).  (top) ALMA image at 241~GHz
  with a peak flux of 0.235~Jy\,beam$^{-1}$. (middle) ALMA image at 97~GHz with a
  peak flux of 0.235~Jy\,beam$^{-1}$. (bottom) VLA image at 43~GHz with a peak flux of
  0.387~Jy\,beam$^{-1}$.  For all images the lowest contour is 0.5\% of the peak
  and the contour levels are in multiplicative increments of
  $\sqrt{2}$. Details of the images are given in Appendix~\ref{sec:3c138}.}
\label{Fig:3C138all}
\end{figure}

The preliminary ALMA 97 \& 241~GHz and VLA 43 GHz images are
shown in Fig.~\ref{Fig:3C138all}.  The bright, compact radio core and strong eastern jet and lobe 
respectively have spectral indices of $-0.70\pm
0.03$ and $-0.75\pm 0.05$. The western counter-jet, which is severely Doppler
attenuated, is weak and has a spectral index of $-0.95\pm 0.13$; its
peak is just below the 3 rms intensity level at 241 GHz. The lowest
contour level for all three images is 0.5\% of their peak intensity (3
times rms), so that the peak to rms ratios for these images are about 500:1.
The main conclusion is that the differences between the ALMA and VLA 
images are at the level of a few percent of their peak levels.  The
two arrays have major differences, such as their antenna, electronics,
and correlator designs; the atmospheric conditions; and ALMA linear
polarized feeds versus the VLA circular polarized feeds.  Hence, the
agreement of the images to a few percent strongly suggests that both
arrays can image the radio emission from the sky at tens of
milliarcsecond resolution with this accuracy or better.

\begin{figure}
\includegraphics[angle=00,width=8cm]{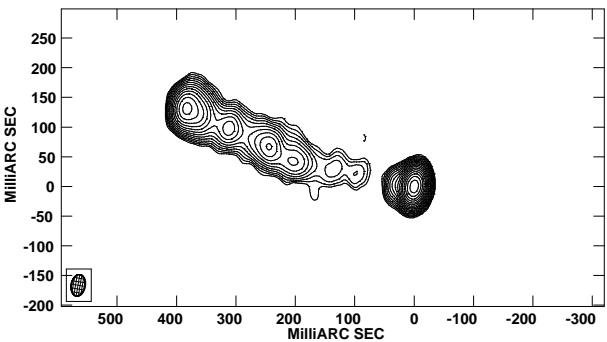}
\caption{Highest resolution ALMA image of 3C138 at 241~GHz. 
The resolution is 37$\times$23 mas at P.A.$=$$-11^\circ$ (shown by the
  cross-hatched ellipse). The contour
  levels are in multiplicative increments of $\sqrt{2}$. The peak flux is 0.095~Jy\,beam$^{-1}$
  and the lowest contour is 0.33\% of the peak.}
\label{Fig:ALMA_Hi}
\end{figure}

The ALMA Band 6 image using the high resolution data at natural weight is
shown in Fig.~\ref{Fig:ALMA_Hi}.  The resolution is 37$\times$23~mas in P.A.
$-11^\circ$ which is considerably higher than that used for the three
frequency comparison.  At this higher resolution, the western jet has broken
into six knots and an inner jet emanating east from the core can be
separated.  The jet/lobe system has a slight
curvature which is also seen on VLBA images of this source \citep{CO2003}.
The faint western counter-jet has a peak flux density of 0.25 mJy,
just below the lowest contour level at 0.3\% of the peak.

\section{Self-Calibration}\label{sec:selfcal}

Some of the SV targets were sufficiently strong that self-calibration could be
used to improve the image quality over that obtained with phase
referencing alone. The Juno images (ALMA Partnership et al. 2015a) were significantly improved with
self-calibration and obtained a peak/image rms of typically 120 for each
of the nine images, providing an increase over the phase-referenced only images of a factor of two to six. 

For the HL Tau continuum images (ALMA Partnership et al. 2015b), self-calibration was more challenging, because while the overall integrated flux is large, the source morphology is complex. Indeed, much of the disk emission is resolved out by the longest baselines, especially at Band 7, and for the lower frequency Bands the emission is intrinsically weaker due to the lower dust emissivity. Thus, the S/N for self-calibration is inadequate for the longest baseline antennas if one attempts to push to short enough timescales ($<$ a few minutes) to significantly improve the phases beyond that achieved from fast-switching. Due to this S/N limitation on the solution interval, the self-calibration only improves the HL~Tau images (peak/rms) by factors of 1.5, 1.9, and 1.2 at Bands 3, 6, and 7 respectively. For the much weaker source SDP.81, there is inadequate S/N to self-calibrate on a short enough timescale to improve the images at all (while retaining the longest baseline antennas). 

Since the 3C138 emission is dominated by a
nearly unresolved core and the remaining structure is relatively
simple, it showed the most improvement.  
The rms noise level decreases about a factor of 10 from the phase
referenced to the self-calibrated image.  A conservative measure is
the ratio of the highest side-lobe level to the peak intensity.  
For the 97 GHz image, the side-lobe/peak intensity ratio drops from 1.4\% in the phase
referenced image to 0.1\% in the self-calibrated image. For
the 241~GHz image, the ratio drops from 17\% to 0.6\%.

One particular complication of self-calibration at long baselines is that unless the target structure
is already well-studied at high resolution, only a rough estimate of
its correlated flux density at the longer baseline may be estimated. Therefore, in many cases it may be difficult to predict in advance whether a given source can be self-calibrated on the longest baselines. 
In future, more sophisticated methods of self-calibration may benefit extended 
sources where the SNR on the longer spacings drops 
below that needed for self-calibration using one reference antenna. 
Furthermore, future testing on long baselines 
will provide further insight into ALMA long-baseline imaging and
self-calibration.

\acknowledgments
This paper makes use of the following ALMA data: ADS/JAO.ALMA\#2011.0.00013.SV, ADS/JAO.ALMA\#2011.0.00014.SV, ADS/JAO.ALMA\#2011.0.00015.SV and ADS/JAO.ALMA\#2011.0.00016.SV. 
ALMA is a partnership of ESO (representing its member states), NSF (USA) and NINS (Japan), together with NRC (Canada), 
NSC and ASIAA (Taiwan), and KASI (Republic of Korea), in cooperation with the Republic of Chile. 
The Joint ALMA Observatory is operated by ESO, AUI/NRAO and NAOJ.
The National Radio Astronomy Observatory is a facility of the National 
Science Foundation operated under cooperative agreement by Associated 
Universities, Inc. 

We thank all those who have contributed to making the ALMA project possible.

{\it Facilities:} \facility{ALMA}.

{}

\end{document}